# Topological defects at octahedral tilting plethora in bi-layered perovskites

F.-T. Huang[1], B. Gao[1], J.-W. Kim[1], X. Luo[2], Y. Wang[1], M.-W. Chu[3], C.-K. Chang[4], H.-S. Sheu[4] and S.-W. Cheong[1,2*]

[1] Rutgers Center for Emergent Materials and Department of Physics and Astronomy, Rutgers University, Piscataway, New Jersey 08854, USA

[2] Laboratory for Pohang Emergent Materials and Max Plank POSTECH Center for Complex Phase Materials, Pohang University of Science and Technology, Pohang 790-784, Korea

[3] Center for Condensed Matter Sciences, National Taiwan University, Taipei 106, Taiwan

[4] National Synchrotron Radiation Research Center, Hsinchu 300, Taiwan

*email: sangc@physics.rutgers.edu

**Oxygen octahedral distortions including tilts/rotations, deformations and off-centering in (layered) perovskites play the key role in their numerous functional properties. Near the polar-centrosymmetric phase boundary in bi-layered perovskite $Ca_{3-x}Sr_xTi_2O_7$ with x≈1, we found the presence of abundant topological $Z_8$ vortex-antivortex pairs, associated with four oxygen octahedral tilts at domains and another four different oxygen octahedral tilts at domain walls. Our discovery opens a new revenue to unveil topological defects associated with various types of oxygen octahedral distortions.**

Copious functional phenomena, including high $T_c$ superconductivity[1], ferroelectricity[2,3], novel magnetism[4-6], and giant photovoltaic effects[7,8], have been observed in perovskite ($ABO_3$)-related compounds, where those physical properties can be closely associated with oxygen octahedral distortions, including tilts/rotations, deformations and off-centering. For example, the high carrier mobility in transparent conducting cubic $BaSnO_3$ (ref. 9) or photovoltaic halide perovskites[8, 10] is closely related with (nearly) 180° bonding between large B-site cations and an oxygen (or a halide ion), resulting from little octahedral distortions. Even when A-site ions are rather small, the perovskite-related structure can be still stabilized through oxygen octahedral tilts/rotations[11]. It turns out that superconductivity in (La,Ba)$_2$CuO$_4$ is significantly influenced by oxygen octahedral tilts[12,13], and canted magnetic moments appear in antiferromagnetic perovskites with tilted/rotated oxygen octahedra through Dzyaloshinskii-Moriya interaction[2, 14]. High dielectric response in the vicinity of the morphotropic phase boundary (MPB) is a consequence of the continuously rotating polarization with various octahedral tilts[15, 16]. Crystal field split can be considerably influenced by compression or elongation of oxygen octahedra, which is the origin of the Jahn-Teller effects for B=$Cu^{2+}$ or $Mn^{3+}$ (ref. 1,17,18).

Remarkably, the simultaneous presence of oxygen octahedral tilt and rotation can result in ferroelectric polarization in perovskites with even number of layers, which is called hybrid improper ferroelectricity (HIF)[19-21]. It has experimentally verified that bi-layered perovskite Ca$_{3-x}$Sr$_x$Ti$_2$O$_7$ (CSTO) is a hybrid improper ferroelectric with switchable polarization of 8 $\mu C \cdot cm^{-2}$ in bulk crystals at room temperature[20]. Ferroelectricity in CSTO described by a hybridization of two structural modes (octahedral tilt and rotation modes) turns out to be associated with an intriguing domain topology consisting of $Z_4 \times Z_2$ domains and $Z_3$ vortices with eight domains (4 directional domains and 2 antiphase domains), abundant charged domain walls and unique zipper-like switching kinetics[22]. In $Z_4 \times Z_2$ domains, $Z_4$ denotes the cyclic group of order 4 for directional variants and $Z_2$ is for translational variants. In this letter, utilizing in-situ heating transmission electron microscopy (TEM) studies, synchrotron powder X-ray diffraction experiments and dielectric measurements, we report the discovery of a new intermediate phase in CSTO, which demonstrates a displacive nature of HIF mechanism different from those predicted from the group-subgroup relation[21,23]. Furthermore, we find the presence of topological $Z_8$ vortex-antivortex defects associated with two-dimensional eight degrees of freedom for oxygen octahedral tilts in the intermediate tetragonal phase.

Fig. 1 depicts the possible octahedral tilts/rotations in bi-layered perovskite CSTO, resulting in different structural phases and also domains with different directional order parameters. For each space group, the distortion-axes and the corresponding Glazer notations[24] are given with respect to the un-distorted tetragonal $I4/mmm$ (T) structure (Fig. 1b). The term "rotation" denotes a rotation of the basal oxygen plane around the $[001]_T$-axis in a clockwise (+) or counterclockwise (-) manner (Fig. 1c for +). Out-of-phase and in-phase rotation in adjacent layers within one bi-layered perovskite block lead to orthorhombic O* and O" phases, respectively. Note that the presence of an intermediate O* phase (Fig. 1c) competing with the ground-state polar O phase, responsible for uniaxial negative thermal expansion, has been confirmed in HIF magnet $Ca_3Mn_2O_7$ (ref. 25). The term "tilt" is associated with a tilt around an in-plane axis, and it would move the basal oxygens out of the basal plane and the apical oxygen away from the c-axis. In-plane tilting axes can be along $<110>_T$ and $<100>_T$-directions, leading to orthorhombic O′ (Fig. 1e) and tetragonal T′ phases (Fig. 1f), respectively. Following the Glazer notation, the T phase has the $a^0a^0c^0$ pattern and the T′, O′, O* and O″ phases are described as $a^-a^0c^0$, $a^-a^-c^0$, $a^0a^0c^-$, and $a^0a^0c^+$, respectively. Two end members, $Sr_3Ti_2O_7$ and $Ca_3Ti_2O_7$, correspond to two extremes: $Sr_3Ti_2O_7$ and $Ca_3Ti_2O_7$ are the un-tilted T phase and the O phase with both tilt and rotation ($a^-a^-c^+$) (Fig. 1d), respectively. It is clear why orthorhombic O′ ($a^-a^-c^0$) and O″ ($a^0a^0c^+$) phases are the two plausible intermediate phases toward the polar O phase ($a^-a^-c^+$) by symmetry[21, 23]. With the underlying square lattice, various symmetry-equivalent domains may exist in each of those phases. It is convenient to define the azimuthal angle of an apical oxygen distortion, $\varphi$, as shown in Fig. 1d. This $\varphi$ links all possible directions of apical oxygen motions among those phases and also domains; for example, symmetry-equivalent domains of the T′ phase correspond to $\varphi$=0°, 90°, 180° and 270° (red-circled ❶, ❷, ❸ and ❹ in Fig. 1f, respectively) while those in the O′ phase to $\varphi$=45°, 135°, 225° and 315° (blue, 1, 2, 3, and 4 in Fig. 1e, respectively); The domains of the polar O phase has $\varphi$=45°±α, 135°±α, 225°±α and 315°±α, where α depends on the sign and magnitude of octahedral rotation ($a^0a^0c^+$) (1±, 2±, 3± and 4± in Fig. 1d, respectively).

We have performed synchrotron powder X-ray diffraction experiments using the collimated synchrotron-radiation beam with the wavelength of 0.688 Å at the National Synchrotron Radiation Research Center, Taiwan. Homogeneous and phase-pure polycrystalline $Ca_{3-x}Sr_xTi_2O_7$ (CSTO, 0≤$x$≤3) specimens were prepared by a solid-state reaction method (Methods). The general structure analysis system (GSAS) program using the Rietveld method with a pseudo-Voigt profile function was exploited to analyze the powder-diffraction data. The evolution of octahedral rotation ($\theta_R$) and tilting ($\theta_T$) angles of $TiO_6$, and lattice parameters $a$, $b$, and $c$ as a function of Sr content, $x$, is summarized in Fig. 2a and Supplementary Fig. S1a. The phase diagram, constructed from these structural parameters, consists of ferroelectric O (purple, 0≤$x$≤0.9) and paraelectric T phases (pink, x≥1.5). The asymmetric decays of $\theta_R$ and $\theta_T$ at $x$=0.915-1 defines a sharp and narrow region with only non-zero $\theta_T$ (yellow, Fig. 2a). The high-resolution x-ray diffraction data clearly display peak splitting's only when $x$≤0.9, implying a tetragonal phase in this narrow region (Supplementary Fig. S1b). A √2×√2 supercell in electron diffraction patterns from transmission electron microscopy (TEM) (Fig. 2c) is found for $x$=0.915-1, confirming the stabilization of a new tetragonal intermediate phase distinct from the parent T phase. The Rietveld refinement of synchrotron data provides further confirmation of the existence of the T′ phase with the $a^-a^0c^0$ pattern (Supplementary Fig. S1c). The details of x-ray refinement fits are given in Supplementary section 1 and Table S1. The dielectric constant also shows a dramatic change in magnitude upon entering the T′ phase and displays almost two times larger epsilon ($\varepsilon$) value at the phase boundary of $x$=0.9, compared to that of low $x$ values (Fig. 2b). An increase in $\varepsilon$ at 350 K can be understood as increasing structural fluctuations when approaching from the ferroelectric O to paraelectric T′ phases. Indeed, in-situ TEM heating experiments of $Ca_{2.1}Sr_{0.9}Ti_2O_7$ crystals exhibit a two-step transformation upon heating: O ➔ T′ ➔ T. Figures 2c shows first that the intensity of superlattice $S_1$-type spots ½(130)$_T$ of the O phase (cyan triangles) weakens as temperature ($T$, defined italic $T$ as temperature) is raised from 300 K to 450 K. Two additional sets of superlattice $S_2$-type ½(200)$_T$ and $S_3$-type ½($\bar{1}$30)$_T$ spots (yellow and green triangles), corresponding to the T′ phase, appear when temperature is further raised to 473 K. Finally, all superlattice spots vanish above 713 K. Thus, we have demonstrated the stabilization of the intermediate T′ phase by varying chemical composition as well as temperature. Starting from the O phase at low $x$, Fig. 2a shows a faster relaxation of the octahedral rotation ($\theta_R$) than that of tilting ($\theta_T$) with increasing Sr doping, which is also coupled with an increasing trend of orthorhombicity[20]. The sudden suppression of octahedral rotations occurs when the azimuthal angle $\varphi$ suddenly switches from ~45° at $x$=0.9 to 0° in the T′ phase (Fig. 1a), indicating a discontinuous change of the tilting order parameter and a first-order phase transition. Note that we do not observe any evidence for an intermediate O′ ($a^-a^-c^0$, $\varphi$=45°) phase, reported in the solid solution of $Ca_3Ti_2O_7$–$(Ca,Sr)_{1.15}Tb_{1.85}Fe_2O_7$ (ref. 26).

Figure 3a and 3b show a 2.7 µm x 1.7 µm mosaic of dark-field (DF) TEM images of the T′-phase $x$=0.95 taken using $S_2$ spot (red circle in Fig. 3b) and its corresponding diffraction pattern along the $[001]_T$ direction. The curved dark-contrast lines reveal boundaries of four T′-phase domains merging at one core, which is a non-T′ phase. With the underlying square lattice, 4 symmetry-equivalent domains, named $Z_1 \times Z_4$ domains, may exist; four translational variants associated with the translation vectors (½, ½, 0), (0, ½, ½) and (½, 0, ½) where out of plane phase shifts are involved in the latter two. However, the domain topology can be renamed as $Z_2 \times Z_2$ domains when the in-plane order parameters (octahedral tilts) of one bilayer is considered; two directional variants ($[010]_T$-tilt producing ❶/❸ and $[100]_T$-tilt producing ❷/❹) and two translational variants associated with the in-pane translation vector (½, ½, 0) between ❶-❸ or ❷-❹ (Fig. 1f). Our results demonstrate that four domains, corresponding to red-circled ❶❷❸❹ in Fig. 1f, form an exclusive vortex-like pattern with 90°–rotating apical oxygen distortions (red arrows in Fig. 3a). These four domains are represented by four values of apical oxygen azimuthal angles $\varphi$=0°, 90°, 180° and 270°; for example, domain ❶ and ❸ correspond to $\varphi$=0° and 180°, respectively. Note that the domain network (Fig. 3a) can be described by two proper coloring, i.e. two colors is sufficient to identify the domains without neighboring domains sharing the same color (white and green in Supplementary Fig. S3a), and the vortex network can be constructed by cutting through two types of closed loops (blue and light blue in Supplementary Fig. S3a) based on a graph-theoretical description[27]. Intriguingly, those curved domain walls exhibit two distinct extinction rules in the superlattice DF-TEM images. Figure 3c shows a DF-TEM image using a $S_1$ spot (light-blue circled in Fig. 3b), revealing only a part of domain walls. A complementary domain-wall map is obtained using a $S_3$ spot for DF-TEM imaging (blue-dotted circled in Fig. 3b). The inequivalent nature of these two DF-TEM images with different superlattice peaks with

*ab*-plane components indicates an orthorhombic-like local structure at those domain walls, instead of, e.g., a high-symmetry T phase. Considering a pure tilting nature of the matrix T′ phase and the distinct extinction rules, those domain walls are likely in the O′-phase state (Fig. 1e). The directions of oxygen octahedral tilts and apical oxygen distortions are denoted as light blue and blue arrows in the Fig. 3c and 3d insets, respectively─ more details are discussed in Supplementary Section 2. Therefore, oxygen octahedra inside the bright contrast domains tilt along either $[100]_T$ or $[010]_T$ directions whereas those in the dark-contrast domain walls tilt along diagonal $<110>_T$ directions. This leads to a unique $Z_8$ vortex structure with four domains and four domain walls, where oxygen octahedra tilt and $\varphi$ changes by 45° consecutively around the vortex core. The full assignment of vortices and antivortices in Fig. 3a is given in Fig. S3a-b, based on the structural information derived from Fig. 3c-d.

Figure 4a summarizes two types of $Z_8$ vortices (types-*i* & *ii*) and two types of $Z_8$ antivortices (types-*iii* & *iv*) observed in Fig. 3a. Topological charge or winding number "*n*" is assigned when vectors rotate clockwise by $2\pi n$ along the clockwise direction around a core. Then, the topological charges of types-*i*, *ii*, *iii*, and *iv* topological defects are +1, +1, -1, and -1, respectively, which indicates that types-*i* & *ii* are vortices and types-*iii* & *iv* are antivortices. We emphasize that vectors in a type-*i* topological defect rotate opposite to those in a type-*ii* topological defect, but they have the same topological charge, so both of them are vortices. A $Z_8$ vortex is always surrounded by antivortices and vice versa; for example, the vortex "*i*" in Fig. 3a is connected to the antivortex "*iii*" and the antivortex "*iv*". Figure 4b illustrates the local structure around a type-*i* $Z_8$ vortex (Fig. 4a) with red-circled ❶, ❷, ❸ and ❹ domains and domain walls of DW❶❷ (blue-broken lines), DW❶❹ (light-blue-broken), DW❸❹ (blue) and DW❷❸ (light-blue). By averaging the structure of neighboring domains, DW❶❷ and DW❸❹ share the same $[110]_T$-tilting axis, but are different in the origin by a half of the unit cell. Similarly, DW❷❸ and DW❶❹ have the same tilting-axis, which is consistent with the extinction rules observed in Fig. 3c-3d. On the other hand, a geometric frustration of oxygen octahedral distortions between two antiphase domains occurs likely in DW❶❸ (Fig. 4c), leading to an undistorted domain wall in a T phase-like high-symmetric state. The absence of $Z_3$ vortices where three types of domain walls merge at one point, reveals the nonexistence of (½, ½, 0)-type APB such as DW❶❸, which, in turn, indicates a much higher energy associated with APBs.

Topological defects in the T′ phase are $Z_8$ vortices in $Z_2 \times Z_2$ domains with different single ($a^-a^0c^0$)-type octahedral tilts. Strain between two adjacent $Z_2 \times Z_2$ domains is naturally accommodated by the two-tilt mode ($a^-a^-c^0$) at the wall. Our observation of $Z_8$ vortex-antivortex networks resulting from two-dimensional eight degrees of freedom provides a unique real-space topology where domain and domain walls are intricately intertwined. Emphasize that the T′ phase has been reported in other bi-layered perovskite magnets such as $Sr_2LnMn_2O_7$ (Ln=Ho, Y)[28] and $Tb_2SrB_2O_7$ (B=Fe, Co) [29-31] and magnetism in (layered) perovskites tends to be strongly coupled with structural distortions, so this unique domain topology may not be limited to structural domains and domain walls in CSTO. We also note that the octahedral distortions in the T′ phase is also analogous to that observed in so called low-temperature tetragonal (LTT) phase of high-$Tc$ superconducting $La_{2-x}Ba_xCuO_4$, in which the $CuO_6$ octahedra also tilt in the $<100>_T$ directions[12,13]. Octahedral distortions are ubiquitous in the plentiful perovskite-family compounds, and our discovery of large-real-space-range self-organized nature of the octahedral tilting modulations provide new insights into understanding the topological nature of domains and domain walls in perovskites with octahedral distortions and its relevance to their physical properties.

**Methods**

Eleven high-quality polycrystalline specimens of $Ca_{3-x}Sr_xTi_2O_7$ ($x$ = 0, 0.3, 0.5, 0.6, 0.8, 0.85, 0.9, 1, 1.5, 2 and 3) were prepared using a solid-state reaction method. Stoichiometric mixtures of $CaCO_3$ (Alfa Aesar 99.95%), $SrCO_3$ (Alfa Aesar 99.99%) and $TiO_2$ (Alfa Aesar Puratronic 99.995%) powders were ground, pelletized and then sintered at 1550-1660 °C for 30h. In the range of $1.1 \leq x < 1.5$, we found a triple-layered $A_4B_3O_{10}$ phase is more stable and favored than the bi-layered $A_3B_2O_7$ phase. The powder specimens for acquiring synchrotron x-ray powder-diffraction data were sealed in 0.2-mm-diameter capillary quartz tubes. All synchrotron x-ray experiments were performed at the National Synchrotron Radiation Research Center, Taiwan. The powder-diffraction patterns were acquired using a collimated synchrotron-radiation beam with the wave length of 0.688 Å (18 KeV) and a beam diameter of ~0.2 mm. The general structure analysis system (GSAS) program using the Rietveld method with a pseudo-Voigt profile function was exploited to analyze the X-ray powder-diffraction (XRD) data. Specimens for transmission electron microscope (TEM) studies were prepared by mechanical polishing, followed by Ar-ion milling, and studied using a JEOL-2010F TEM. We observed $Z_8$ vortex domains with superlattice DF-TEM imaging taking (1) $S_1=½(130)_T=(120)_{orth}$, (2) $S_2=½(020)_T=(110)_{orth}$, and (3) $S_3=½(\bar{1}30)_T=(210)_{orth}$ spots. In-situ heating TEM experiment was carried out using a JEOL-2000FX TEM with a high-temperature specimen holder. For dielectric constant measurements, two electrodes were prepared using Au sputtering on polished specimens with a capacitor geometry, and an LCR meter at f = 44 kHz was utilized.

**Acknowledgements**

The work at Rutgers was funded by the Gordon and Betty Moore Foundation's EPiQS Initiative through Grant GBMF4413 to the Rutgers Center for Emergent Materials, and the work at Postech was supported by the Max Planck POSTECH/KOREA Research Initiative Program [Grant No. 2011-0031558] through NRF of Korea funded by MSIP.

**Competing financial interests:** The authors declare no competing financial interests

## Author contributions

F.-T. H. and M.-W. C. conducted the TEM experiments. B. G. and X. L. synthesized single crystals. J.W.K. and Y.W performed the dielectric measurments. C.-K. C and H.-S. S. acquired synchroton x-ray data. F.T-H and S.-W.C wrote the paper. S.-W.C. initiated and supervised the research.

## Additional information

The authors declare no competing financial interests. Supplementary information is available at npj Quantum Materials website. Reprints and permissions information is available online at xxx.

**Data availability.** The authors declare that all source data supporting the findings of this study are available within the article and the Supplementary Information file.

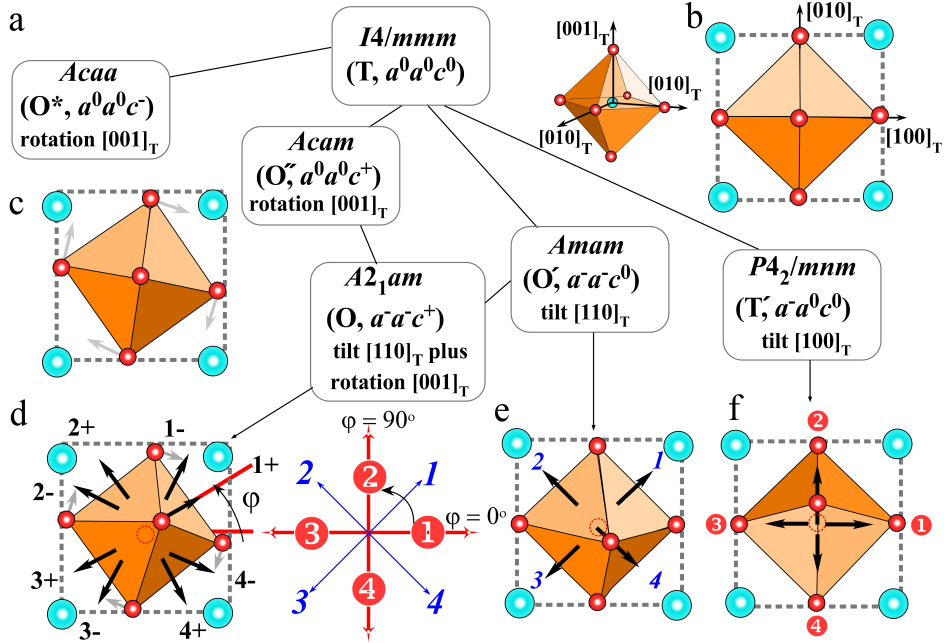

**FIG. 1. a,** Schematic diagram of possible space groups and the corresponding Glazer notations for bi-layered perovskite $A_3B_2O_7$. Lines link the group-subgroup relations. The lattice directions are given with respect to the T phase (*I4/mmm*). **b-f,** In-plane projected views of various $BO_6$ octahedral tilts/rotations. The dashed lines outline the primitive unit cells, and the red and cyan spheres represent the O and B-site ions, respectively. Black and gray arrows indicate directions of apical and basal plane oxygen displacements, respectively. **b,** Un-distorted $BO_6$ octahedron in the T phase. **c,** A clockwise (+) rotated $BO_6$ in the O* and O′ phases. The rotation can be also counterclockwise (−). In two adjacent layers within one bi-layer, an out-of-phase rotation (+ − or − +) along the c-axis occurs in the O* phase while in-phase (+ + or − −) rotation in the O″ phase. **d,** Eight possible 1±, 2±, 3±, and 4± apical oxygen displacements in the O phase. The azimuthal angle, $\varphi$, of an apical oxygen distortion is denoted. **e,** Given tilting vectors in the O′ phase. **f,** Four possible apical oxygen displacements (red-circled labels) along the $<100>_T$ directions in the T′ phase.

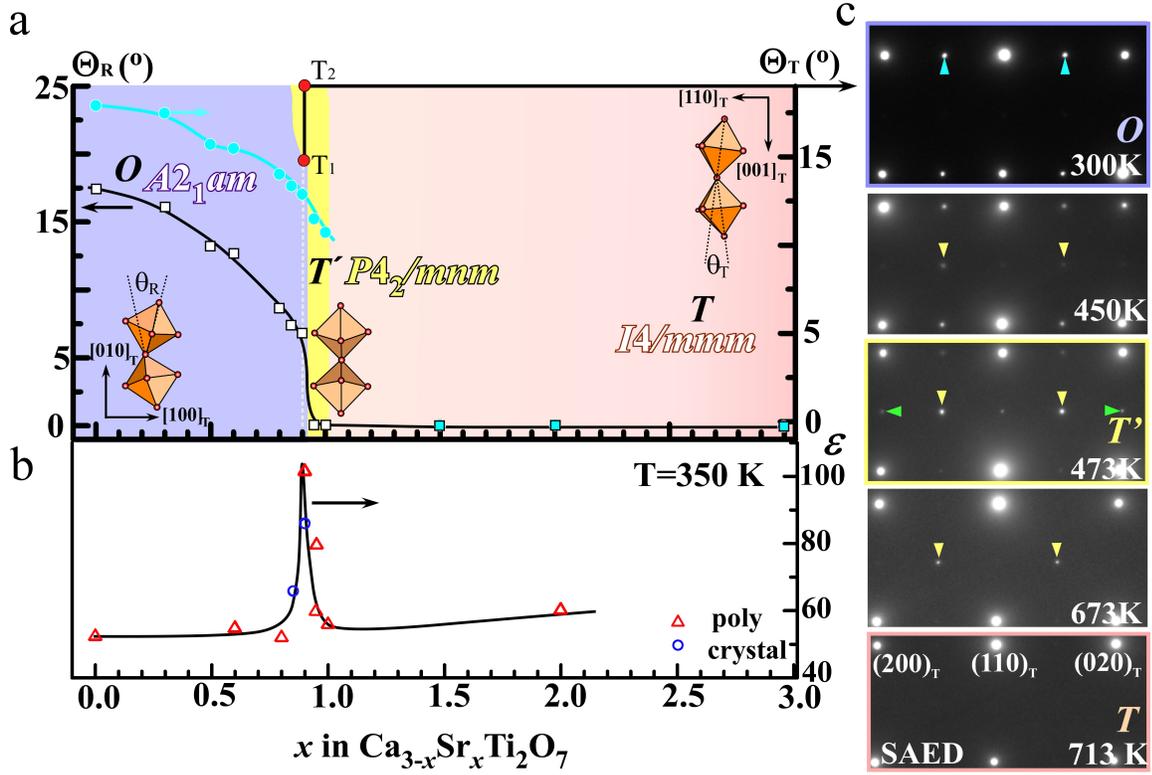

**FIG. 2. a,** Structural phase diagram of $Ca_{3-x}Sr_xTi_2O_7$ with the angles of octahedral rotation ($\theta_R$, empty squares) and tilt ($\theta_T$, cyan spheres) as a function of Sr doping, $x$. The black and cyan curves are guides for the eyes. Two structural phase transition temperatures ($T_1$=473 K and $T_2$=710 K) at $x$=0.9, marked with red spheres, are obtained from in-situ TEM heating experiments. **b,** Dielectric constant vs. Sr content ($x$), measured at 350 K and 44 kHz. Red triangles and blue circles are from measurements on polycrystalline and single-crystalline (electric field applied along the in-plane direction) specimens, respectively. The dielectric constant peaks at the O-T′ phase boundary of $x$=0.9. **c,** Thermal sequence of selected area electron diffraction patterns (SAED) on $x$=0.9, showing two phase transitions upon in-situ heating: O (300 K) ➔ T′ (473 K)➔ T (713 K) phases. $S_1$=½(130)$_T$-type superlattice spots (cyan triangles) are allowed in the O phase. When temperatures reaches 473 K, additional $S_2$=½(200)$_T$-type and $S_3$=½(-130)$_T$ type superlattice spots appear (yellow and green triangles), indicating the appearance of a tetragonal T′ phase. Above 713 K, all superlattice spots disappear, consistent with the presence of the T phase at very high temperatures.

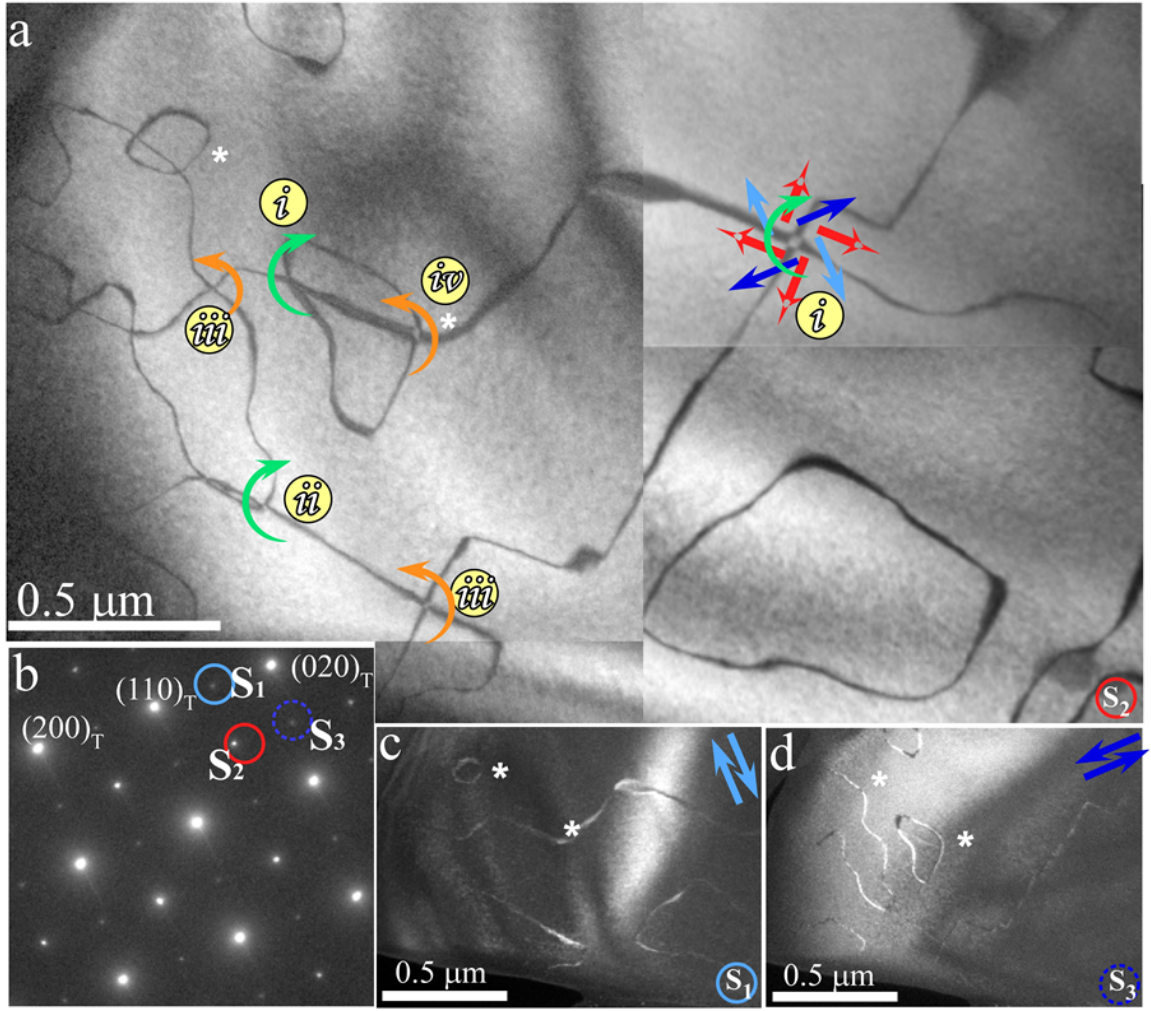

**FIG. 3. a,** A 2.7 μm x 1.7 μm mosaic of DF-TEM images, taken using a superlattice $S_2$ spot (red-circled) in a CSTO $x$=0.95 specimen along the $[001]_T$ direction. The colored arrows represent octahedral tilting directions; red arrows for four domains of the T′ phase and (light)-blue arrows for four domain walls in the O′-phase state. Yellow-circled *i-iv* are $Z_8$ vortex-antivortex defects. Asterisks are location markers. **b,** The corresponding SAED. **c,** A superlattice DF-TEM image, taken using a superlattice $S_1$ spot (light blue-circled), exhibits only a part of domain walls in (a). **d,** A superlattice DF-TEM image, taken using a $S_3$ spot (blue-dot circled), shows the rest of domain walls. Octahedral tilting directions in real space are shown in the up-right corner.

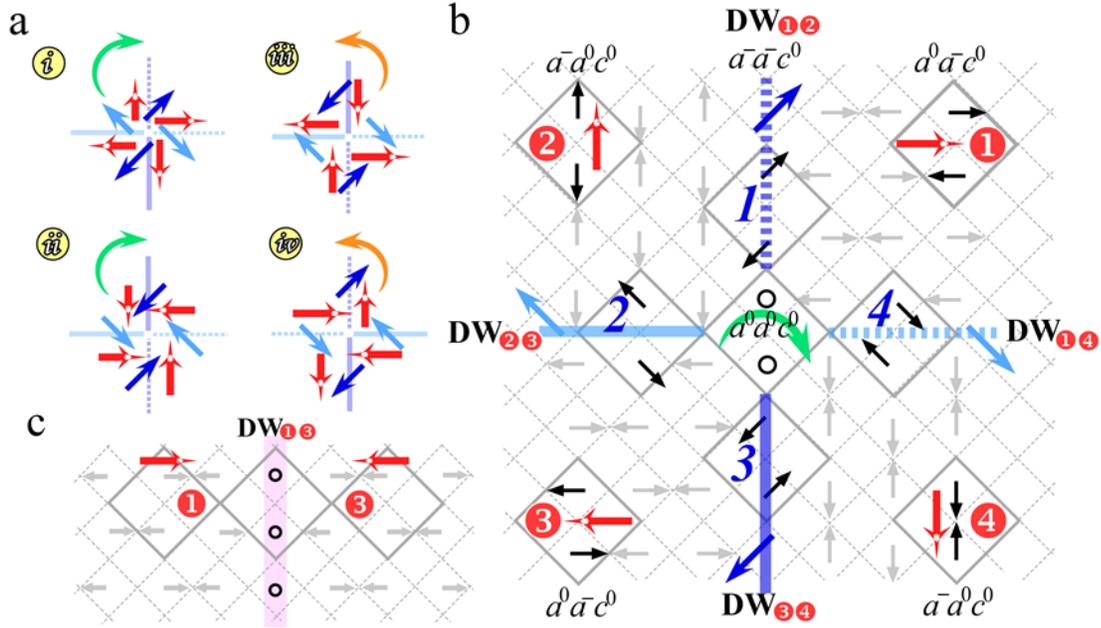

**FIG. 4. a,** Schematics illustrate the oxygen octahedral distortion configurations at domains and domain walls of $Z_8$ vortex-antivortex defects shown in Fig. 3a. Red arrows for four domains show apical oxygen tilts along the $<100>_T$ directions. Blue and light-blue bold lines represent two types of domain walls originated from two different tilting axes, and solid and broken lines are related by a translational symmetry. Yellow-circled types-*i* & *ii* are $Z_8$ vortices and type-*iii* & *iv* are $Z_8$ antivortices. **b,** Local oxygen distortions of a type-*i* $Z_8$ vortex with four domains (red-circled ❶-❹) and four domain walls (blue 1-4) in the *ab*-plane. The local apical oxygen distortions in each domain and domain wall are shown by small gray and black arrows. The gray solid squares outline the √2×√2 supercell with the T′ ($P4_2/mnm$) symmetry inside domains and with the O′ (*Amam*) symmetry at domain walls. A frustration of oxygen octahedral tilts among four domains suggests an un-distorted structure (open circles) for the $Z_8$ vortex core. The Glazer notations label the dominant tilting modes at the corresponding locations. **c,** A schematic of the domain wall between two antiphase domains, *e.g.* DW❶❸ between domain ❶ and ❸. The open circles indicate the ideal position of oxygens of the T phase.

17